\documentclass[twocolumn,showpacs,aps,prd,amsmath,amssymb,nofootinbib,nobibnotes,floatfix]{revtex4-1}
\usepackage{hyperref}
\usepackage{amsfonts}
\usepackage{amsmath}
\usepackage{mathrsfs}
\usepackage{epsfig}
\usepackage{graphicx}
\usepackage{url}
\usepackage{color,soul}
\usepackage{hyperref}
\usepackage{float}
\usepackage{color}
\usepackage[utf8]{inputenc} 
\usepackage{graphicx}
\usepackage{epsf}
\usepackage{comment}
\usepackage{slashed}
\usepackage{footmisc}
\usepackage{setspace}
\usepackage[dvipsnames, usenames]{xcolor}
\definecolor{linkcolor}{rgb}{0.0,0.3,0.5}

\usepackage{hyperref}
\hypersetup{
    pdfnewwindow=true,      
    colorlinks=true,       
    linkcolor=linkcolor,          
    citecolor=linkcolor,        
    filecolor=blue,      
    urlcolor=blue           
}

\begin{document}

\title{Probing the Black Hole Merger History in Clusters using Stellar Tidal Disruptions}

\author{Johan Samsing$^{1}$, Tejaswi Venumadhav$^{2}$, Liang Dai$^{2}$, Irvin Martinez$^{3}$, Aldo Batta$^{3}$, Martin Lopez Jr.$^{4}$, Enrico Ramirez-Ruiz$^{3,4}$, Kyle Kremer$^5$}
\affiliation{\vspace{2mm}
$^1$Department of Astrophysical Sciences, Princeton University, Peyton Hall, 4 Ivy Lane, Princeton, NJ 08544, USA.\\
$^2$School of Natural Sciences, Institute for Advanced Study, 1 Einstein Drive, Princeton, New Jersey 08540, USA\\
$^3$Niels Bohr Institute, Blegdamsvej 17, 2100 Copenhagen, Denmark\\
$^4$Department of Astronomy and Astrophysics, University of California, Santa Cruz, CA 95064, USA\\
$^5$Center for Interdisciplinary Exploration and Research in Astrophysics (CIERA) and Department of Physics \& Astronomy, Northwestern University, Evanston, IL 60208, USA
}

\begin{abstract}
The dynamical assembly of binary black holes (BBHs) in dense star clusters (SCs) is one of the most promising
pathways for producing observable gravitational wave (GW) sources, however several other formation scenarios likely operate as well. One of the current outstanding questions is how these different pathways may be distinguished apart. In this paper we suggest a new multi-messenger observable that can be used to constrain the formation of BBH mergers originating from SCs: the electromagnetic signal from tidal disruptions (TDs) of stars by BBHs. Such TDs will show variability in their light curve from the orbital motion of the disruptive BBHs, and can therefore be used to map the BBH orbital period distribution, and thereby also the dynamical mechanisms that eventually drive the BBHs to merger. Using an analytical approach including General Relativistic effects, we find that the orbital period distribution of BBHs within globular clusters peaks on timescales of days, which we argue is unique to this assembly pathway. We propose that the search for variable TDs in current and future EM transient surveys might be used to constrain the merger history of BBHs in SCs.
\end{abstract}

\maketitle

\section{Introduction}\label{sec:Introduction}
During the first and second observing runs of the advanced gravitational-wave detector network, LIGO and VIRGO detected ten merging binary black holes (BBHs) \citep{2016PhRvL.116f1102A, 2016PhRvL.116x1103A, 2016PhRvX...6d1015A,
2017PhRvL.118v1101A, 2017PhRvL.119n1101A, 2018arXiv181112907T}. Upgraded detectors are expected to detect hundreds of such mergers within the next few years \citep[see  e.g.,][]{2016ApJ...833L...1A}. The origin and distribution of these BBHs is a key scientific question that remains to be addressed \citep[e.g.,][]{2018arXiv180605195B}. 

Several formation channels have been suggested in the literature: dense stellar clusters \citep{2000ApJ...528L..17P,
2010MNRAS.402..371B, 2013MNRAS.435.1358T, 2014MNRAS.440.2714B,
2015PhRvL.115e1101R, 2016PhRvD..93h4029R, 2016ApJ...824L...8R,
2016ApJ...824L...8R, 2017MNRAS.464L..36A, 2017MNRAS.469.4665P},
field binaries \citep{2012ApJ...759...52D, 2013ApJ...779...72D, 2015ApJ...806..263D, 2016ApJ...819..108B,
2016Natur.534..512B, 2017ApJ...836...39S, 2017ApJ...845..173M, 2018ApJ...863....7R, 2018ApJ...862L...3S},
active galactic nuclei discs \citep{2017ApJ...835..165B,  2017MNRAS.464..946S, 2017arXiv170207818M},
galactic nuclei \citep{2009MNRAS.395.2127O, 2015MNRAS.448..754H,
2016ApJ...828...77V, 2016ApJ...831..187A, 2016MNRAS.460.3494S, 2017arXiv170609896H},
single-single GW captures of primordial black holes (BHs) \citep{2016PhRvL.116t1301B, 2016PhRvD..94h4013C,
2016PhRvL.117f1101S, 2016PhRvD..94h3504C},
and very massive stellar mergers \citep{Loeb:2016, Woosley:2016, Janiuk+2017, DOrazioLoeb:2017}.
There are significant uncertainties in the predicted merger rates in these scenarios due to the complexity of the underlying astrophysical environments. The mergers observed by LIGO are in the gravitational-wave dominated regime, and are hence described only by a few intrinsic parameters, such as the masses and the spins of the black holes. Given a population of detected mergers, the distributions of these intrinsic parameters can be used to disentangle these channels. 

Binary black holes formed in dynamical environments, such as globular clusters (GCs), galactic nuclei (GN),
and nuclei star clusters (NSCs), can be eccentric in both LIGO \citep{2009MNRAS.395.2127O, Kocsis:2012ja, 2014ApJ...784...71S, 2016ApJ...824L..12O, 2017ApJ...840L..14S, 2018MNRAS.476.1548S, 2018ApJ...853..140S, 2019MNRAS.482...30S, 2018PhRvD..97j3014S, 2018ApJ...855..124S, 2018arXiv181000901Z, 2018arXiv181104926R, 2018ApJ...860....5G} and LISA bands \citep[e.g.,][]{2018MNRAS.tmp.2223S, 2018arXiv181111812K}, and have randomly oriented spins when exchange interactions are effective \citep[e.g.,][]{2016ApJ...832L...2R}. In comparison, isolated field binaries are likely to have correlated spins
and merge on near circular orbits \citep[e.g.][]{2018ApJ...862L...3S}.
Exceptions to this standard picture include field BBHs undergoing Lidov-Kozai (LK, \citep[e.g.,][]{1962AJ.....67..591K}) oscillations that tend to re-orient spins \citep[e.g.,][]{2017ApJ...846L..11L, 2018MNRAS.480L..58A, 2018ApJ...863...68L} and produce eccentric mergers \citep[e.g.,][]{2017ApJ...836...39S, 2018arXiv180506458A}, and dynamically formed BBHs that have their spins re-oriented and possibly aligned
through mass accretion following tidal disruptions in dense clusters \citep[e.g.][]{2018arXiv181201118L}. Binaries of primordial BHs that are formed via GW-driven single-single capture are also expected to show signs of eccentricity \citep[e.g.,][]{2016PhRvD..94h4013C}. As a non-zero eccentricity would indicate a
dynamical origin of BBH mergers, major effort is therefore being undertaken to observationally probe eccentric mergers \citep[e.g.,][]{2017PhRvD..95b4038H, 2018PhRvD..97b4031H, 2018arXiv180502716H, 2018arXiv180900672G}.
Various methods have been suggested for distinguishing between dynamical channels, including correlating with the BBH masses, cosmic merger evolution \citep[e.g.,][]{2018ApJ...866L...5R}, and using
future multi-band GW observations \citep[e.g.,][]{2017ApJ...842L...2C, 2018MNRAS.tmp.2223S, 2018MNRAS.481.4775D}.

In this paper, we discuss a new electromagnetic window into the assembly of BBH mergers in dense stellar clusters (SCs): tidal disruptions (TDs) of stars by BBHs. Previous theoretical work on these events suggests that their lightcurves are interrupted on characteristic timescales related to the orbital period of the disrupting BBHs \citep[e.g.,][]{Liu:2009fl, 2014ApJ...786..103L, 2016MNRAS.458.1712R, 2017MNRAS.465.3840C, 2018arXiv181201118L}. We propose that populations of such events can be used to probe the orbital period distribution of BBHs inside SCs.
We explore in detail how the dynamics that drive the BBHs to merger \citep{2006ApJ...637..937O, 2018PhRvL.120o1101R, 2018PhRvD..97j3014S, 2018MNRAS.tmp.2223S, 2018arXiv181104926R} also shape their orbital period distribution. We use these insights to derive, for the first time, a simple analytical model of the BBH period distribution with the inclusion of General Relativistic (GR) effects,
and relate the TD and BBH merger rates from SCs.

We find that the orbital period distribution of BBHs within standard GCs
peaks on timescales of days, with only a weak dependence on BH mass.
This period distribution is entirely different from other theoretical distributions derived under simple considerations, such as those driven purely by GW emission \citep{2017MNRAS.469..930C}, and the empirical Opik's Law distribution for field binaries  \cite{1924PTarO..25f...1O}.
This emboldens us to suggest that if current and future transient surveys observe a population of interrupted TDs, the inferred orbital periods will directly inform us about the dynamical assembly of BBHs in dense SCs. Provided we properly account for the observational biases of the surveys and understand the factors governing the detectability of the TDs, such detections can also constrain the contribution of SCs to the BBH merger rate in the universe. 

The paper is structured as follows. In Section \ref{sec:Stellar Disruptions by Binary Black Holes}
we describe our GR dynamical model and derive an analytical expression for the rate distribution of
star-BBH TDs as a function of the BBH orbital time. This is followed by Section \ref{sec:Limits on the Binary Black Hole Orbital Period}, where we discuss
different astrophysical limits on the BBH orbital distribution. In Section \ref{sec:Results} we summarize our main results
from our dynamical models, where in Section \ref{sec:comp_MC} we compare to numerical data derived
using Monte Carlo (MC) techniques.
In Section \ref{sec:Discussion} we discuss prospects and challenges
related to using interrupted TDs to probe the BBH orbital distribution. We conclude our study in Section \ref{sec:Conclusions}.

\section{Stellar Disruptions by Binary Black Holes}\label{sec:Stellar Disruptions by Binary Black Holes}

In this section we derive the rate of stars disrupted by BBHs in SCs per BBH
orbital period, as a function of the BH masses and SC properties.
For this, we use our analytical model presented in \cite{2018PhRvD..97j3014S, 2018MNRAS.tmp.2223S} for describing
the GR dynamical evolution of the BBHs, that we throughout this work assume are equal mass. This is a reasonable approximation
as both mass-segregation and few-body dynamics tend to keep equal mass objects together \citep[e.g.,][]{Heggie:1975uy, 2016PhRvD..93h4029R, 2017MNRAS.469.4665P}.
Our derivation is divided into a few separate steps, as indicated by the following subsections. Below we start
by describing our dynamical model.

\subsection{Dynamical Model}\label{sec:Dynamical Model}

\begin{figure}
\centering
\includegraphics[width=\columnwidth]{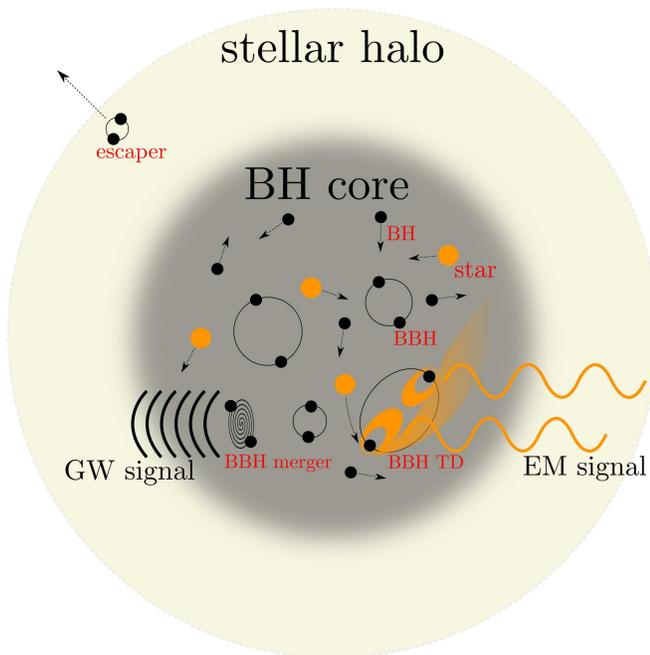}
\caption{Illustration of a stellar cluster composed of a `stellar halo' and a `BH core'.
In the core the number density of BHs occasionally reaches values that facilitate the formation of BBHs through
random three-body encounters \citep[e.g.,][]{Heggie:1975uy, 1976A&A....53..259A, Hut:1983js}. These BBHs then
undergo binary-single interactions with incoming objects, particularly including other BHs (`BH') and stars (`star').
Each binary-single interaction with other similar mass BHs leads to a decrease in the BBH's SMA by a factor $\delta \approx 7/9$ \citep[e.g.,][]{2018PhRvD..97j3014S},
as well as resets its eccentricity according
to a thermal distribution $P(e) = 2e$ \cite[e.g.,][]{Heggie:1975uy}. The decreasing SMA and the possibility for entering a high
eccentricity state makes it possible for the BBH to undergo a GW merger inside the cluster in-between its binary-single interactions (`BBH merger').
If the BBH does not undergo a GW merger in-between or during its interactions \citep[e.g.,][]{2018MNRAS.tmp.2223S}, it will instead
get dynamically ejected and escape the cluster (`escaper').
Since BBHs with a relative small SMA $a$ are likely to merge inside the cluster due to their corresponding short GW life time ($\propto a^{4}$), one expects to see
relative few BBHs with short orbital periods ($T \propto a^{3/2}$). Therefore, if the period distribution of BBHs can be
mapped, one will be able to indirectly constrain the past and current BBH merger history.
In this paper we suggest that the BBH orbital period distribution can be sampled using interrupted stellar tidal disruptions (`BBH TD').
This serves as a new `multi-messenger probe' that connects LIGO signals (`GW signal') with electromagnetic observations (`EM signal').
}
\label{fig:GCill}
\end{figure}

In our model we assume that BBHs form at a steady rate in the SC core, through the interaction of three
initially unbound BHs \citep[e.g.,][]{Heggie:1975uy, 1976A&A....53..259A, Hut:1983js}, with their semi-major axes (SMA) equal
to the hard-binary (HB) value \citep{Hut:1983js},
\begin{equation} 
a_{\rm HB} = \frac{3}{2}\frac{ G m_{\rm BH}}{v_{\rm dis}^{2}},
\label{eq:aHB}
\end{equation}
where $m_{\rm BH}$ denotes the mass of one of the (assumed equal mass) BHs, and $v_{\rm dis}$ is the SC velocity
dispersion. After formation the dynamics of the BBHs is governed by binary-single interactions involving other (equal mass) BHs.
Each of these interactions leads to a decrease in the SMA of the interacting BBH from SMA $a$ to $\delta a$, a process referred to as `hardening',
where we assume a constant value $\delta = 7/9$, which is the mean value for equal mass interactions \citep{2018PhRvD..97j3014S}.
Following this approximation, a given BBH will therefore after $n$ interactions have a SMA equal to
\begin{equation}
a_{\rm n} = a_{\rm HB} \times \delta^{n}.
\label{eq:defan}
\end{equation}
In this paper we will occasionally refer to a BBH with SMA $a_{\rm n}$ to be in `state $n$'  or at `hardening step $n$'.
This binary-single hardening process continues until either the release of binding energy from a binary-single interaction
recoils the BBH out the cluster \citep[e.g.,][]{2000ApJ...528L..17P}, or that the interacting BBH undergoes a GW merger
{\it during} \citep[e.g.,][]{2006ApJ...640..156G, 2014ApJ...784...71S, 2018arXiv181000901Z, 2018arXiv181104926R}
or {\it in-between} \citep[e.g.,][]{2018PhRvL.120o1101R, 2018PhRvD..97j3014S, 2018arXiv181104926R}
its binary-single interactions. For further descriptions of this model we refer the reader to \cite{2018MNRAS.tmp.2223S}.
Information is also provided in Figure \ref{fig:GCill}. 

\subsection{Formulation}

We start by considering an ensemble of SCs each with a constant production rate
of BBHs that undergo binary-single hardening according to our model described in the above Section \ref{sec:Dynamical Model}.
We consider this distribution of BBHs at a random point
in time -- the time of observation -- at which we observe a set of stars disrupted by the BBHs. The question is, what is the distribution
of BBH orbital periods for the set of BBHs disrupting the stars?
For answering this, we start by writing the differential tidal disruption rate per $\log$ BBH orbital time as a product of the following terms,
\begin{equation}
\frac{d\Gamma_{\rm TD}}{d\log{T}} =  \frac{dN_{\rm BBH}}{dn}  \frac{d\Gamma_{\rm TD}}{dN_{\rm BBH}}  \frac{dn}{d\log{T}},
\label{eq:dGdlogT}
\end{equation}
where $T$ is the BBH orbital period, $\Gamma_{\rm TD}$ is the rate of stars disrupted by BBHs,
$n$ is the $n$'th hardening step, and $N_{\rm BBH}$ is the number of BBHs present at state $n$ at the time of observation.
The last term ${dn}/{d\log{T}}$ is a simple Jacobian factor that can be calculated from first using that $T_{\rm n} \propto a_{\rm n}^{3/2} \propto \delta^{3n/2}$, where
$T_{\rm n}$ denotes the BBH orbital period at state $n$, which follows from Kepler's laws
and Eq. \eqref{eq:defan}. This implies that $\log T_{\rm n} \propto n(3/2) \log \delta \propto n$, from which we conclude that ${dn}/{d\log{T}}$
simply equals a constant. In the following subsections we derive the two remaining
terms ${dN_{\rm BBH}}/{dn}$ and ${d\Gamma_{\rm TD}}/{dN_{\rm BBH}}$,
where our final expression for ${d\Gamma_{\rm TD}}/{d\log{T}}$ is presented in Section \ref{sec:Distribution of BBH-Stellar Disruptions}.

\subsection{Distribution of Binary Black Holes}

For calculating the term ${dN_{\rm BBH}}/{dn}$, i.e. the number of BBHs per state $n$ at the time of observation,
we start by factorizing it as,
\begin{equation}
\frac{dN_{\rm BBH}}{dn} \propto \mathscr{W}_{n} \times \mathscr{P}(n),
\label{eq:dNBHdn}
\end{equation}
where $\mathscr{W}_{n}$ is a weight factor that essentially equals the probability that the BBH is in state $n$ at the time of observation given that
it is not able to undergo any type of merger inside the cluster (Newtonian term), and $\mathscr{P}(n)$ is the probability that the BBH do not
merge before and during state $n$ (GR term). Note that we do not consider the merging population in our calculations, i.e. we only
account for the BBHs in Eq. \eqref{eq:dNBHdn} that also `survives' state $n$, although there is a small probability
for the merging BBHs to disrupt stars during their GW inspiral: This represents an interesting case, but is beyond what we can study with our
semi-analytical models. In the following two sections we derive the `observational weight
factor' $\mathscr{W}_{n}$ and the term describing the `depletion from GW Mergers' $\mathscr{P}(n)$, respectively.

\subsubsection{Observational Weight Factor}

The factor $\mathscr{W}_{n}$ is proportional to the time the BBH spends at state $n$,
which in our model equals the time between binary-single interactions at state $n$, a time we denote $t_{\rm bs,n}$.
The factor $\mathscr{W}_{n}$ can therefore be approximated by \citep[e.g.,][]{2018arXiv180708864S},
\begin{equation}
\mathscr{W}_{n} \propto t_{\rm bs,n} \approx \left({\eta_{\rm BH} \sigma_{\rm bs,n} v_{\rm dis}}\right)^{-1} \propto \delta^{-n},
\label{eq:Pobsn}
\end{equation}
where $\eta_{\rm BH}$ is the number density of single BHs in the SC core,
and $\sigma_{\rm bs,n} \propto m_{\rm BH}a_{\rm n}/v_{\rm dis}^{2}$
is the cross section for a strong binary-single interaction at state $n$ \citep[e.g.,][]{2018ApJ...853..140S}.

\subsubsection{Depletion from Gravitational Wave Mergers}

\begin{figure}
\centering
\includegraphics[width=\columnwidth]{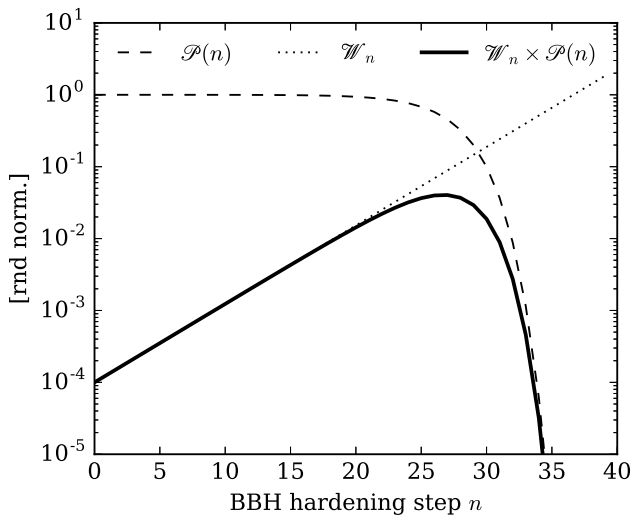}
\caption{The probability that a BBH is present at hardening step $n$ at the time of observation,
is the product of two terms: the probability that the BBH do not merge before and during state
$n$, $\mathscr{P}(n)$ (Eq. \eqref{eq:Pnfin}), and the probability that we observe it if it is in state $n$, $\mathscr{W}_{n}$ (Eq. \eqref{eq:Pobsn}).
These two terms are shown in the above figure with {\it dashed}
and {\it dotted} lines, respectively, where their product $dN_{\rm BBH}/dn \propto \mathscr{W}_{n} \times \mathscr{P}(n)$ is shown with a
{\it solid} line. For this figure we have assumed $m_{\rm BH} = 30M_{\odot} $, $v_{\rm dis} = 15\ \text{kms}^{-1}$,
and $\eta_{\rm BH} = 10^{5}\ \text{pc}^{-3}$.
As seen, the term $\mathscr{P}(n)$ takes the form of a super-exponential decay near $n \approx 30$. When GR effects are included BBHs are therefore less likely to be observable
at larger $n$, which corresponds to lower SMA $a_{\rm n}$ or equivalently lower BBH orbital time $T_{\rm n}$. The peak
seen in $dN_{\rm BBH}/dn$ maps to a characteristic BBH orbital period that is unique to stellar clusters as we show in Fig. \ref{fig:dGdlogT}.
}
\label{fig:PnPobs}
\end{figure}

To derive $\mathscr{P}(n)$ we first express it as the following product,
\begin{equation}
\mathscr{P}(n)= \prod_{i=0}^{n} \mathscr{P}_{i} = \prod_{i=0}^{n} (1 - \tilde{\mathscr{P}_{i}}),
\end{equation}
where $\mathscr{P}_{i}$ denotes the probability for that the BBH do not merge at state $i$,
and $\tilde{\mathscr{P}_{i}}$ that the BBH do merge at state $i$. To evaluate this product
we first rewrite it as,
\begin{equation}
\ln \mathscr{P}(n)= \sum_{i=0}^{n} \ln (1 - \tilde{\mathscr{P}_{i}}) \approx - \sum_{i=0}^{n} \tilde{\mathscr{P}_{i}},
\label{eq:lnP1}
\end{equation}
where for the last term we have assumed that $\tilde{\mathscr{P}_{i}} \ll 1$.
The term $\tilde{\mathscr{P}_{i}}$ denotes the total probability for the BBH to merge at state $i$, which in our model
either can happen in the time-span before it undergoes its next binary-single
interaction \citep[e.g.,][]{2018MNRAS.tmp.2223S} or during the binary-single
interaction itself through a two-body GW capture \citep{2014ApJ...784...71S}.
Normally, the probability for undergoing a merger
in-between binary-single interactions is far larger than doing a binary-single interaction, therefore,
to leading order $\tilde{\mathscr{P}_{i}}$ is well approximated by the probability that the BBH at state $i$ has a GW inspiral
life time, $t_{{\rm m},i}$, less than its binary-single encounter time, $t_{{\rm bs},i}$.
Using that the GW inspiral life time can be written as $t_{{\rm m},i} \approx t_{{\rm cm},i}\times(1-e^{2})^{7/2}$, where $t_{{\rm cm},i}$
denotes the circular GW inspiral life time for which the BBH eccentricity $e = 0$ \citep{Peters:1964bc}, and assuming that the BBH eccentricity
distribution follows that of a so-called thermal distribution $P(e) = 2e$ \citep{Heggie:1975uy}, the probability $\tilde{\mathscr{P}_{i}}$ can then
be written as \citep[e.g.,][]{2018ApJ...853..140S, 2018PhRvD..97j3014S},
\begin{equation}
\tilde{\mathscr{P}_{i}} \approx  \left(t_{{\rm bs},i}/t_{{\rm cm},i} \right)^{2/7} = \tilde{\mathscr{P}}_{\rm HB} \times {\varepsilon}^{i}, 
\label{eq:tilPi}
\end{equation}
where $\varepsilon \equiv \delta^{-10/7}$, and $\tilde{\mathscr{P}}_{\rm HB}$ is the probability that the BBH merges before its next binary-single interaction
at the HB limit when its SMA $=a_{\rm HB}$. This probability can be expressed as,
\begin{equation}
\tilde{\mathscr{P}}_{\rm HB} \approx  \left( \frac{8192}{3645\pi} \frac{1}{G^{3}c^{5}} \frac{v_{\rm dis}^{11}}{m_{\rm BH}^{3}\eta_{\rm BH}} \right)^{2/7},
\end{equation} 
which follows directly from Eq. \eqref{eq:Pobsn}, Eq. \eqref{eq:tilPi}, and the relation $t_{{\rm cm},i} \propto a_i^{4}/m_{\rm BH}^{3}$
given by \cite{Peters:1964bc}. Now using the relation from the above Eq. \eqref{eq:tilPi} we can write Eq. \eqref{eq:lnP1} as,
\begin{equation}
\ln \mathscr{P}(n) \approx - \tilde{\mathscr{P}}_{\rm HB} \sum_{i=0}^{n} {\varepsilon}^{i}.
\end{equation}
This sum can be evaluated analytically using the solution $\sum_{i = 0}^{n}x^{i} = (1-x^{n+1})/(1-x)$, which follows from the standard
literature on `geometrical series'. With this, we finally find,
\begin{equation}
\mathscr{P}(n) \approx \exp \left(- \tilde{\mathscr{P}}_{\rm HB} \times \frac{1-{\varepsilon}^{n+1}}{1-{\varepsilon}} \right).
\label{eq:Pnfin}
\end{equation}
As seen, the effect from including the possibility for a BBH to merge in-between its binary-single interactions gives rise to a BBH
depletion that takes the form of a super-exponential decay. We note here that the BBHs that merge in-between their
binary-single interactions, i.e. the population giving rise to the decay, have been shown to constitute $\approx 30-50\%$ of the observable
BBH merger rate from GCs \citep[e.g.,][]{2018PhRvL.120o1101R, 2018MNRAS.tmp.2223S, 2018arXiv181104926R}; the function $\mathscr{P}(n)$ can
therefore be directly linked to LIGO/Virgo events.
One should note that the above expressions are only valid for $n<n_{\rm max}$, where $n_{\rm max}$ is the value of $n$ for which
the corresponding $\tilde{\mathscr{P}_{n}} = 1$. Using Eq. \eqref{eq:tilPi} we find,
\begin{equation}
n_{\rm max} \approx -\frac{\log \tilde{\mathscr{P}}_{\rm HB}}{\log {\varepsilon}}.
\label{eq:nmax}
\end{equation}
Fig. \ref{fig:PnPobs} illustrates the terms $\mathscr{W}_{n}$, $\mathscr{P}(n)$, and the factor
${dN_{\rm BBH}}/{dn} \propto \mathscr{W}_{n} \times \mathscr{P}(n)$ given by Eq. \eqref{eq:dNBHdn}, as a function
of $n$ for values describing a dense SC system (see figure caption). The maximum value $n_{\rm max}$ for these values is $n_{\rm max} \approx 31$, which 
indeed is where $\mathscr{P}(n)$ starts to rapidly fall off. The limit for $n_{\rm max}$ is not naturally build into the equations,
as we assumed in Eq. \eqref{eq:lnP1} that $\tilde{\mathscr{P}_{i}} \ll 1$, which helped us evaluating the sum.
Our main results are not affected by these small differences. Other upper limits on $n$ exists as well, some of which will
be described in Section \ref{sec:Limits on the Binary Black Hole Orbital Period}.

Finally, we note that in \cite{2018PhRvD..97j3014S} the total probability for a BBH to merge in-between its binary-single interactions during hardening
from the HB limit to step $n$ was derived in the integral-limit using slightly different approximations (Section III.D.2 in \cite{2018PhRvD..97j3014S}), from
which it was found that $\tilde{\mathscr{P}}(n) \approx (7/10)\tilde{\mathscr{P}}_{n}/(1-\delta)$. In certain limits our derived expression from Eq. \eqref{eq:Pnfin}
should be identical to that previous result. After some mathematical manipulations, one indeed finds that
$\tilde{\mathscr{P}}(n) = 1-\exp (- \tilde{\mathscr{P}}_{\rm HB} {(1-{\varepsilon}^{n+1})}/{(1-{\varepsilon})} )$ is $= (7/10)\tilde{\mathscr{P}}_{n}/(1-\delta)$ in the limit where
$\mathscr{P}(n) \gg 0$, $\tilde{\mathscr{P}}_{n} \gg \tilde{\mathscr{P}}_{\rm HB}$, and $\delta \approx 1$, all of which were assumed in \cite{2018PhRvD..97j3014S}.
This serves as an excellent check of the different approaches that so far have been employed to estimate the probability for a BBH to undergo a GW merger
during hardening.

\subsection{Rate of Tidal Disruptions}\label{sec:Rate of Tidal Disruptions}

To estimate the term ${d\Gamma_{\rm TD}}/{dN_{\rm BBH}}$, i.e. the stellar tidal disruption rate per BBH, we first factorize it as,
\begin{equation}
\frac{d\Gamma_{\rm TD}}{dN_{\rm BBH}} \approx \frac{d\Gamma_{\rm s}}{dN_{\rm BBH}} \times P_{\rm TD},
\end{equation}
where $\Gamma_{\rm s}$ denotes the rate of star-BBH interactions, and $P_{\rm TD}$ the probability for that a single star-BBH interaction
results in a stellar tidal disruption.
The first term $d\Gamma_{\rm s}/dN_{\rm BBH}$ is $\propto \eta_{\rm s} \sigma_{\rm s} v_{\rm dis}$, where $\eta_{\rm s}$ is the number density of
stars, and $\sigma_{\rm s} \propto m_{\rm BH}a/v_{\rm dis}^2$ is the cross section for a star-BBH interaction \citep[e.g.,][]{2017ApJ...846...36S}.
The second term $P_{\rm TD}$ is proportional to the probability for that the incoming star undergoes a pericenter passage w.r.t.
to one of the two BHs during the star-BBH interaction that is $\lesssim R_{\rm TD}$, where $R_{\rm TD}$ is the stellar disruption distance.
In the equal mass case estimating this probability is straight forward and can e.g. be done by describing the temporary three-body state as a binary with
a bound single \citep[e.g.,][]{2018PhRvD..97j3014S}; however, the relative small mass of the star in our case will rarely allow it to form a bound system
with any of the BHs. Simulations show that the light star instead will undergo chaotic single encounters with
each BH on orbits with a wide range of energies and eccentricities \citep[e.g.,][]{2017MNRAS.465.3840C, 2018MNRAS.477.4009D, 2018ApJ...863L..24C, 2018arXiv181201118L}.
Especially, one finds that the star is highly likely to disrupt on orbits that, relative to the disrupting BH, are {unbound}, which is possible since each BH
evolves with a relative high velocity around their common center-of-mass \citep[e.g.][]{2018arXiv181201118L}.
Although one might be able to develop an analytical description of this limit, we here make use of the recent work by \cite{2018MNRAS.477.4009D},
which indicates the probability $P_{\rm TD}$ is $\propto R_{\rm TD}/a$ (we note that this exact same scaling is found in
the equal mass case \citep[e.g.,][]{2017ApJ...846...36S, 2018PhRvD..97j3014S}). Using this scaling one now finds, 
\begin{equation}
\frac{d\Gamma_{\rm TD}}{dN_{\rm BBH}} \propto {\eta_{\rm s} \sigma_{\rm s} v_{\rm dis}} \times \frac{R_{\rm TD}}{a}  \propto \frac{\eta_{\rm s} m_{\rm BH} R_{\rm TD}}{v_{\rm dis}}.
\end{equation}
From this one concludes that ${d\Gamma_{\rm TD}}/{dN_{\rm BBH}}$ is independent of the BBH
orbital period $T$ for fixed $m_{\rm BH}$.
The term ${d\Gamma_{\rm TD}}/{dN_{\rm BBH}}$ acts therefore only as a normalization factor, and
plays therefore no role in the calculation of the functional form for ${d\Gamma_{\rm TD}}/{d \log T}$.
Finally, we note that one do expect corrections to this simple picture when the stellar radius becomes comparable to the BBH SMA $a$ \citep[e.g.][]{2018arXiv181201118L}.

\subsection{Final Solution}\label{sec:Distribution of BBH-Stellar Disruptions}

Using that $\delta^{n} \propto T_{\rm n}^{2/3}$ one can now finally write the solution to Eq. \eqref{eq:dGdlogT} in terms of $T$ as,
\begin{equation}
\begin{split}
\frac{d\Gamma_{\rm TD}}{{d\log{T}}} & \propto \frac{dN_{\rm BBH}}{dn} \propto \frac{dN_{\rm BBH}}{d\log T}\\
			& \propto T^{-2/3} \exp{\left(-\tilde{\mathscr{P}}_{\rm HB} \times \frac{1-{\varepsilon}(T/T_{\rm HB})^{-20/21}}{1-{\varepsilon}}\right)},
\end{split}
\label{eq:dGdlogTfinalT}
\end{equation}
where $T_{\rm HB}$ denotes the BBH orbital period at the HB limit,
\begin{equation}
T_{\rm HB} =  \frac{\pi \sqrt{27}}{2}\frac{Gm_{\rm BH}}{v_{\rm dis}^{3}},
\label{eq:THB}
\end{equation}
which follows from Eq. \eqref{eq:aHB}. Note here that in the above Eq. \eqref{eq:dGdlogTfinalT} we have further
indicated that the derived distribution is proportional to the term ${dN_{\rm BBH}}/{d\log T}$.
This is a highly useful quantity that will be relevant for other and future studies. One should
also note that the above Eq. \eqref{eq:dGdlogTfinalT} does not include possible lower limits on $T$,
but this will be discussed in further detail in Section \ref{sec:Limits on the Binary Black Hole Orbital Period},
before we explore our solution in Section \ref{sec:Results}.

Finally, to check our derivations, we compared our expression
from Eq. \eqref{eq:dGdlogTfinalT} with results derived using our
semi-analytical MC code described in \cite{2018MNRAS.tmp.2223S}. We find excellent agreement, which serve as a good validation of our results so far. In Section \ref{sec:comp_MC} we show comparison of our models to data derived using
a H\'{e}non-type MC code.

\section{Limits on the Binary Black Hole Orbital Period}\label{sec:Limits on the Binary Black Hole Orbital Period}

The upper limit on the BBH orbital period $T$, denoted by $T_{\rm max}$, is in our model
always set by the HB value given by Eq. \eqref{eq:THB}; however,
different dynamical and astrophysical effects will determine the lower limit, $T_{\rm min}$.
In the following we discuss the three limits: `Time Limit' (${T_{\rm min,\tau}}$), `Merger Limit' ($T_{\rm min,m}$), and `Ejection Limit' (${T_{\rm min,ej}}$), so that $T_{\rm min} = \max([T_{\rm min,\tau}, T_{\rm min,m}, T_{\rm min,ej}])$.
In the sections below we quote numbers based on the following `fiducial values':
$m_{\rm BH} = 30M_{\odot} $, $v_{\rm dis} = 12\ \text{kms}^{-1}$, $v_{\rm esc}/v_{\rm dis} = 5$, and
$\eta_{\rm BH} = 10^{4}\ \text{pc}^{-3}$.

\subsubsection{Time Limit}
It takes a finite time $\tau$ for a BBH to harden from its initial orbital time $T_{\rm HB}$ to some smaller
time $T_{\tau}$. An imposed limit on $\tau$ will therefore map to a corresponding lower value on $T$, that we here
denote $T_{\rm min,\tau}$. The absolute upper limit
on $\tau$ is the Hubble time $t_{\rm H}$. For deriving $T_{\rm min,\tau}$, we start by writing out the time it takes for a BBH to reach
state $n$, $\tau_{\rm n} = \sum_{i=0}^{i=n-1} t_{\rm bs, i} = t_{\rm bs, HB} (1-(T_{\rm n}/T_{\rm HB})^{-2/3})/(1-1/\delta)$, where
we have used that $t_{{\rm bs}, i} = t_{\rm bs, HB} \times \delta^{-i}$ which follows from Eq. \eqref{eq:Pobsn},
and $t_{\rm bs, HB}$ refers to the time between
binary-single interactions at the HB limit. This expression can be rearranged from which we now find,
\begin{equation}
\begin{split}
T_{\rm min,\tau}	& \approx {T_{\rm HB}} \times \left(t_{\rm bs, HB}/\tau \right)^{3/2} \left(\frac{\delta}{1-\delta}\right)^{3/2} \\
				& \propto m_{\rm BH}^{-2} v_{\rm dis}^{3/2} {\eta_{\rm BH}^{-3/2}},
\end{split}
\end{equation}
where $T_{\rm min,\tau} \approx 0.3\ \text{days}$ for our chosen fiducial values and $\tau = t_{\rm H}$.
Note here the strong dependence on $m_{\rm BH}$.

\subsubsection{Merger Limit}
\begin{figure}
\centering
\includegraphics[width=\columnwidth]{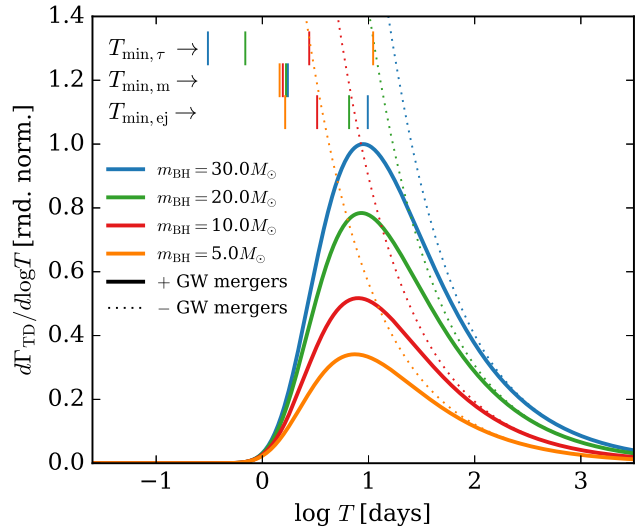}
\caption{Steady-state rate of stellar tidal disruptions due to BBHs as a function of the orbital period of the binaries, derived for our `fiducial values' $v_{\rm dis} = 12\ \text{kms}^{-1}$, $\eta_{\rm BH} = 10^{4}\ \text{pc}^{-3}$,
for various values of the black hole mass $m_{\rm BH}$. The {\it solid lines} show our GR solution from Eq. \eqref{eq:dGdlogTfinalT} that takes into account that BBHs
can merge in-between their hardening binary-single interactions ($+$ GW mergers), whereas the {\it dotted lines} show the
Newtonian solution $\propto T^{-2/3}$ ($-$ GW mergers).
The peak of the distribution depends only weakly on $m_{\rm BH}$ (see Eq.~\eqref{eq:Tpeak}).
The vertical solid lines in the upper part of the figure show
reasonable values of the lower cutoff of the distribution, set by the Hubble time (top-row), and the probabilities of merger and dynamical ejection (center and bottom rows), respectively. See Section \ref{sec:Limits on the Binary Black Hole Orbital Period} for details.
}
\label{fig:dGdlogT}
\end{figure}

In our model it is impossible for a BBH to pass state $n_{\rm max}$ given by Eq. \eqref{eq:nmax} through binary-single interactions alone.
Since $T_{\rm n} \propto (9/7)^{-3n/2}$ then the upper limit $n_{\rm max}$ directly maps to a lower limit on $T$, referred to as $T_{\rm min,m}$.
This limit can be found by the use of Eq. \eqref{eq:nmax}, or by solving for the period $T$ that satisfies $t_{\rm bs} =t_{\rm cm}$, from which follows,
\begin{equation}
\begin{split}
T_{\rm min,m}	 	& \approx {T_{\rm HB}} \times \tilde{\mathscr{P}}_{\rm HB}^{21/20} \\
				& \propto {m_{\rm BH}^{1/10}v_{\rm dis}^{3/10}}{\eta_{\rm BH}^{-3/10}},
\end{split}
\label{eq:Tmin}
\end{equation}
where $T_{\rm min,m} \approx 1.7\ \text{days}$ for our chosen fiducial values. Note that this limit is not automatically included in Eq. \eqref{eq:dGdlogTfinalT}
as we assumed that $\tilde{\mathscr{P}_{i}} \ll 1$ in Eq. \eqref{eq:lnP1} to evaluate the summation; however, Eq. \eqref{eq:dGdlogTfinalT} 
does rapidly decline just around that limit as we later illustrate in Fig. \ref{fig:dGdlogT}.

\subsubsection{Ejection Limit}
A BBH undergoing binary-single interactions inside a stellar cluster receives in each interaction (if no merger takes place)
a dynamical velocity kick $v_{\rm kick}$ that increases with increasing $n$, i.e. decreasing SMA $a_{\rm n}$ \citep[e.g.,][]{1992ApJ...389..527H, 2018PhRvD..97j3014S}.
The BBH in its hardening process will therefore eventually receive a $v_{\rm kick}$ that exceeds the escape velocity $v_{\rm esc}$
of the cluster. As shown in \cite{2018PhRvD..97j3014S}, the value of the BBH SMA for which $v_{\rm kick} = v_{\rm esc}$ can be
written as $a_{\rm ej} = ({1}/{6})\left({1}/{\delta} - 1\right){m_{\rm BH}}/{v_{\rm esc}^2}$.
This marks a lower limit on the BBH SMA, as no BBH is able to pass $a_{\rm ej}$ without being ejected from the cluster (see Fig. \ref{fig:GCill}).
The corresponding minimum orbital period $T_{\rm min, ej}$ is from Kepler's laws given by,
\begin{equation}
\begin{split}
T_{\rm min,ej}	& \approx {T_{\rm HB}} \times (v_{\rm esc}/v_{\rm dis})^{-3} \left( \frac{1-\delta}{9\delta} \right)^{3/2} \\
			& \propto m_{\rm BH} v_{\rm dis}^{-3},
\end{split}
\label{eq:Tminej}
\end{equation}
where $T_{\rm min,ej} \approx 9.8\ \text{days}$ for our chosen fiducial values. Note here the strong
dependence on $v_{\rm dis}$.
Comparing with the other two considered limits,
$T_{\rm min,ej}$ takes the largest value and represents therefore the actual lower limit for our fiducial values.
However, our three considered limits scale differently with $m_{\rm BH}, v_{\rm dis}, \eta_{\rm BH}$,
and the two other limits might therefore dominate in other cases. We comment on this below,
where we discuss our results including the derived limits.

\section{Results}\label{sec:Results}

The distribution ${d\Gamma_{\rm TD}}/{d\log{T}}$ given by Eq. \eqref{eq:dGdlogTfinalT} is plotted in
Fig. \ref{fig:dGdlogT} as a function of log $T$ for individual BH masses $m_{\rm BH} [M_{\odot}] = 5,10,20,30$.
The {\it solid lines} show our derived solution that includes the possibility for GW merger during hardening, where the {\it dotted lines} show the Newtonian
solution $\propto T^{-2/3}$ obtained from setting $\tilde{\mathscr{P}}_{\rm HB} = 0$ in Eq. \eqref{eq:dGdlogTfinalT}.
As seen, the four distributions all peak near the same orbital time $T$.
Furthermore, the peak locations are neither located near the minimum ($T_{\rm min}$) nor the maximum orbital time ($T_{\rm HB}$),
but somewhere in between. The explanation follows from the two terms in Eq. \eqref{eq:dNBHdn}: BBHs with high values of $T$ are less likely to be
observed due to their relative small $t_{\rm bs}$ (Eq. \eqref{eq:Pobsn}), where the number of
BBHs with low values of $T$ is greatly reduced due to depletion through
mergers during the hardening (Eq. \eqref{eq:Pnfin}). This is in contrast to the Newtonian solution, which
predicts that the $T$ distributions always will peak near their lowest possible value. 

To see how the location of the peak value, $T_{\rm peak}$, scales with $m_{\rm BH}$ and the cluster parameters,
we can solve for the peak location using the standard procedure by differentiating ${d\Gamma_{\rm TD}}/{d\log{T}}$ w.r.t. $\log T$,
equal the expression to zero, and then solve for $\log T$.
From this we find that $T_{\rm peak}$ can be written as,
\begin{equation}
\begin{split}
T_{\rm peak}	& = {T_{\rm HB}} \times \tilde{\mathscr{P}}_{\rm HB}^{21/20} \left( \frac{10}{7}\frac{{\varepsilon}}{{\varepsilon}-1} \right)^{21/20} \\
			& \propto {m_{\rm BH}^{1/10}v_{\rm dis}^{3/10}}{\eta_{\rm BH}^{-3/10}}
\end{split}
\label{eq:Tpeak}
\end{equation}
where $T_{\rm peak} \approx 8.9\ \text{days}$ for our chosen fiducial values.
As seen, the location is indeed almost independent of $m_{\rm BH}$ due to the suppression $1/10$ in the exponent,
and only weakly dependent on $v_{\rm dis}$ and $\eta_{\rm BH}$.

In Fig. \ref{fig:dGdlogT} we have shown with vertical lines the
lower limits $T_{\rm min,t_{\rm H}}$, $T_{\rm min,m}$, and $T_{\rm min,ej}$, defined in
Section \ref{sec:Limits on the Binary Black Hole Orbital Period}.
As seen, for our chosen values, $T_{\rm min}$ for BH masses
$m_{\rm BH} [M_{\odot}] = 5,10,20,30$ is $T_{\rm min,\tau}(>T_{\rm peak})$,
$T_{\rm min,ej}(<T_{\rm peak})$, $T_{\rm min,ej}(<T_{\rm peak})$, $T_{\rm min,ej}(>T_{\rm peak})$, respectively.
It is clear that $T_{\rm min,ej}$ is likely to be the most relevant limit for merging BBHs in classical GCs;
however, the value of $T_{\rm min,ej}$ depends strongly on $v_{\rm dis}$ as seen in Eq. \eqref{eq:Tminej}. This implies,
for example, that for our chosen cluster values a dispersion $v_{\rm dis} \gtrsim 20\ \text{kms}^{-1}$ the lower limit $T_{\rm min}$ will instead be $T_{\rm min,m}$ for all our considered values of $m_{\rm BH}$. In that case ${T_{\rm peak}}/{T_{\rm min,m}} =  \left(({10}/{7}){{\varepsilon}}/({{\varepsilon}-1})\right)^{21/20} \approx 5$, i.e.,
the peak $T_{\rm peak}$ is always visible in the high velocity dispersion limit, and no sources are
expected to be found with $T \lesssim T_{\rm peak}/5$. This limit further implies and confirms that in NSCs all
BBHs will merge inside the system before dynamical ejection takes place \citep[e.g.,][]{1995MNRAS.272..605L, 2016ApJ...831..187A}, which greatly increases the possibility for repeated mergers and possible build-up of heavy central objects \citep[e.g.,][]{2016ApJ...831..187A, 2018arXiv181103640A}.

\section{Comparing to MC Simulations}\label{sec:comp_MC}

\begin{figure}
\centering
\includegraphics[width=\columnwidth]{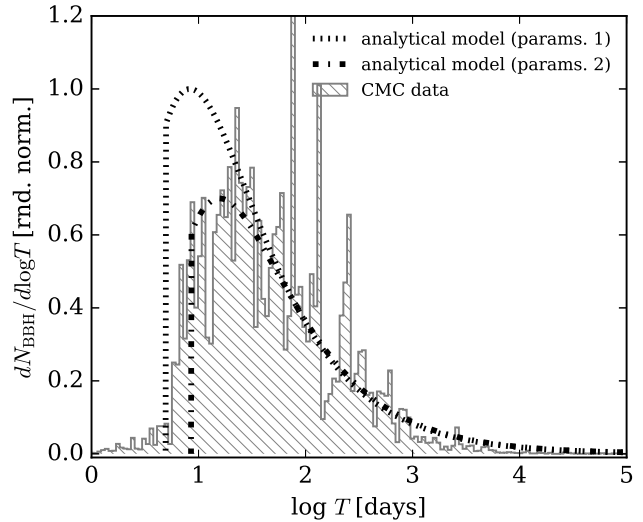}
\caption{Number of BBHs with orbital period $T$, as a function of $\log{T}$,
evaluated at a specific snap-shot in time (the time of observation) for a population of SCs.
The {\it grey-hatched} histogram `CMC data' shows data from the H\'{e}non-type MC code \texttt{CMC},
extracted from a population of GCs all with an initial number of particles $N=2 \times 10^6$ sampled
in time according to their initial metalicity, as further described in Section \ref{sec:comp_MC}.
Only BBHs where each BH has a mass $10M_{\odot}< m_{\rm BH} <20M_{\odot}$
has been included to provide a fair comparison with our equal-mass models.
The black {\it dotted} and {\it dash-dotted} lines show our analytical solution from
Eq. \eqref{eq:dGdlogTfinalT}, using two different set of cluster values
`params. 1' ($m_{\rm BH} = 15M_{\odot} $, $v_{\rm dis} = 12\ \text{kms}^{-1}$, $v_{\rm esc}/v_{\rm dis} = 5$,
$\eta_{\rm BH} = 10^{4}\ \text{pc}^{-3}$), and
`params. 2' ($m_{\rm BH} = 15M_{\odot} $, $v_{\rm dis} = 10\ \text{kms}^{-1}$, $v_{\rm esc}/v_{\rm dis} = 5$,
$\eta_{\rm BH} = 10^{3}\ \text{pc}^{-3}$), respectively. Despite our simplified assumptions compared to
the highly complex astrophysical and dynamical effects taking into account by the \texttt{CMC} code,
the two approaches show excellent overlap, which validates our presented analytical framework. Further discussions are given in Section \ref{sec:comp_MC}.
}
\label{fig:dNBBHdlogT}
\end{figure}

In Figure \ref{fig:dNBBHdlogT} we compare our analytical derived solutions
from Section \ref{sec:Distribution of BBH-Stellar Disruptions} to the BBH orbital period distribution, $dN_{\rm BBH}/d\log T$, extracted from full-scale GC models developed using the MC code, \texttt{CMC}. \texttt{CMC} is a H\'{e}non-type MC
code used to model the long-term evolution of GCs, incorporating various evolutionary effects including two-body relaxation, single and binary-star evolution, and small-$N$ resonant encounters with GR corrections (for a review and summary of the latest updates to \texttt{CMC}, see \citep{Henon1971a,Henon1971b,Joshi2000,Joshi2001,Fregeau2003,Pattabiraman2013,Chatterjee2010,Chatterjee2013,Rodriguez2015a,Rodriguez2018a,Rodriguez2018b}).
The data we compare to is taken from the \texttt{CMC} models used in \citep{2018arXiv181111812K} and \citep{Rodriguez2018b}. We consider only those \texttt{CMC} models with $N=2 \times 10^6$ particles, which most closely match our chosen fiducial cluster parameters ($v_{\rm dis} = 12\ \text{kms}^{-1}$, $v_{\rm esc}/v_{\rm dis} = 5$, and
$\eta_{\rm BH} = 10^{4}\ \text{pc}^{-3}$; see Section \ref{sec:Limits on the Binary Black Hole Orbital Period}). In order to generate a GC population representative of the clusters found in the local universe at present, we use a scheme similar to that of \citep{2018arXiv181111812K} where the cluster models are assigned weights based on the present-day GC mass function and metallicity distributions of \citep{El-Badry2018}. Additionally, cluster ages are drawn from the metallicity-dependent age distributions of \citep{El-Badry2018}. We perform $10^4$ weighted cluster draws from the \texttt{CMC} models and then extract the orbital parameters of all BBHs found in each cluster draw at the present-day in order to generate the \texttt{CMC} orbital period distribution in Figure \ref{fig:dNBBHdlogT}. The two black curves in the figure show our solution given
by Eq. \eqref{eq:dGdlogTfinalT}, with lower cutoffs set by $T_{\rm min,ej}$ derived using Eq. \eqref{eq:Tminej},
for two different set of cluster parameters (`params. 1' and `params. 2'), as specified in the caption.

The overlap between the \texttt{CMC} data and our analytical solution is surprisingly good.
The distribution $dN_{\rm BBH}/d\log T$ derived from \texttt{CMC} follows exactly our main predictions
including an initial rise $\propto T^{-2/3}$ with a lower cutoff set by a combination of
dynamical ejections and relativistic BBH merger depletion. As seen, the `params. 2' values seem to fit the data slightly
better than our fiducial values. It would be very interesting to explore how a MC derived
distribution $dN_{\rm BBH}/d\log T$ changes with a wider range of GC parameters, metallicities, and age; however,
that is out of the scope of this paper.
With this excellent agreement between numerical and analytical
techniques, we proceed below by discussing our results.

\section{Discussion}\label{sec:Discussion}

In the previous sections, we derived the distribution of the orbital periods of BBHs in SCs.
Here we start by comparing this result to the distributions expected in other astrophysical environments. We finish by discussing the prospects for and challenges involved in using interrupted TDs as probes of the BBH orbital time distribution, and comment on other probes.

\subsection{Orbital Distributions}

The orbital period distribution of BBHs in SCs is shaped by the same GR dynamics that also drive the BBHs to merger. A measure of the orbital period distribution can therefore be used to indirectly probe the corresponding BBH merger history.
However, this is only possible if the distribution from SCs is different from those found in other scenarios. In the paragraphs below we compare our derived distribution given by Eq. \eqref{eq:dGdlogTfinalT},
\begin{equation}
\frac{dN_{\rm BBH}}{d \log T} \propto T^{-2/3} \mathscr{P}(T)\ \ \text{(dynamics)},
\end{equation}
to BBHs following Opik's Law and BBHs that evolve through GW emission only.

The `Opik's Law distribution' suggests that the distribution of binary SMA $a$ 
follows $dN_{\rm BBH}/da \propto 1/a$ in the field \cite{1924PTarO..25f...1O}. This relation corresponds to an orbital distribution scaling as,
\begin{equation}
\frac{dN_{\rm BBH}}{d \log T} \propto \text{const.} \ \ \text{(Opik's Law)}.
\end{equation}
Opik's distribution corresponds therefore to a flat distribution in $\log T$. This is clearly different from our derived SC distribution which decreases for increasing $\log T$.

For BBHs evolving by GW radiation only,
the steady state solution in the $e=0$ limit can be found from solving a homogeneous advection equation \citep{2016PhRvL.116w1102S, 2017MNRAS.469..930C}.
The solution to this is $dN_{\rm BBH}/dT \propto T^{5/3}$, which corresponds to,
\begin{equation}
\frac{dN_{\rm BBH}}{d \log T} \propto T^{8/3} \ \ \text{($e = 0$, GW driven)}.
\end{equation}
A population of BBHs entirely driven by GW radiation is therefore expected
to have its period distribution increasing with increasing $\log T$. This behavior is opposite to what is found in our dynamical
driven SC case.

To conclude, our derived $dN_{\rm BBH}/d \log T$ distribution of BBHs in SCs
is clearly different from what is expected of BBHs not driven by dynamics.
Therefore, if a distribution is measured that favors short period BBHs
with a peak $T_{\rm peak} \sim$ days, then this would suggest that
BBHs in SCs are present in the nearby universe and contribute to the observed GW
merger rate. However, since the limit $T_{\rm min, ej}$ depends
strongly on $v_{\rm dis}$, it is not entirely clear what the observed average value and dispersion is
of the lower limit $T_{\rm min}$ for a representative GC ensemble. More detailed GC simulations
must be used to provide further insight into this.

\subsection{Probing the BBH Orbital Time}

Several observational probes exist that can be used to infer the orbital time distribution of
BBHs in SCs. In the sections below we start by discussing prospects and challenges for using EM signals following
a stellar TD event, after which we sketch out a few other possibilities.

\subsubsection{Stellar Tidal Disruptions}

A star disrupted by a BBH is likely to give rise to a EM signal that initially follows the standard $t^{-5/3}$ decline \citep{Rees1988, 2013ApJ...767...25G}, after which the orbital motion of the two BHs gives rise to
interruptions after a time $t_{\rm itr}$ that is comparable to the BBH orbital time $T$. This interestingly suggests that one is able to infer $T$ from observations of the interruption
time $t_{\rm itr}$ \citep[e.g.,][]{Liu:2009fl, 2014ApJ...786..103L, 2018arXiv181201118L}.
This approach was applied in \cite{2014ApJ...786..103L}, who studied the TDE J1201+30 which started out
as a canonical TD after which clear interruptions appeared
$\approx 27$ days after discovery. Using a simple $N$-body code
for resolving the stream dynamics \citep{Liu:2009fl}, the authors
estimated the orbital time and eccentricity of the disrupting
super-massive BBH (SMBBH) to be $\approx 150$ days and $\approx 0.3$, respectively.
The scenario of star-BBH TDs has been studied for both TDs by SMBBHs
($m_{\rm BH} \sim 10^{7}M_{\odot}$) \citep[e.g.][]{Liu:2009fl, 2016MNRAS.458.1712R, 2017MNRAS.465.3840C, 2018MNRAS.477.4009D} and recently by stellar mass BBHs ($m_{\rm BH} \sim 15M_{\odot}$) \citep{2018arXiv181201118L}, and generally it is found that the rate of
accretion indeed is linked to the BBH's orbital parameters. However, it is far
from clear to what degree the EM light-curve actually traces the rate of mass fallback
$\dot{M}$, which often is taken as a proxy for the EM emission from the TDE \citep{2018arXiv180110180S}.
If the emission is beamed through a relativistic jet, one also expects light-curve variations
from both precession and orbital motion caused by the companion BH and possible also the spin of the disrupting BH \citep[e.g.][]{2011Natur.476..421B, 2013PhRvD..87h4053S}.
As further discussed in \cite{2018arXiv181201118L} and briefly below, we are
likely only to observe jetted TDEs, and the jet dynamics for a disrupting stellar
mass BBHs is therefore important to understand. This is a non-trivial problem and we therefore
keep that for a future study.

We now turn to the astrophysical rates of star-BBH TDs expected from GCs. This rate depends both on the number density of stars, the absolute number of BBHs, but not on the the distribution
of SMA a of the BBHs (see Section \ref{sec:Rate of Tidal Disruptions}). A simple `$n \sigma v$' estimate gives,
\begin{equation}
\frac{\Gamma_{\rm TD}^{\rm BBH}}{\text{gal.}} \approx 10^{-7}\ \text{yr}^{-1}\ \left(\frac{\eta_{\rm s}}{10^{4}\text{pc}^{-3}}\right) \left(\frac{m_{\rm BH}}{30M_{\odot}}\right)^{4/3}\left(\frac{12\text{kms}^{-1}}{v_{\rm dis}}\right),
\label{eq:GTDEBBH}
\end{equation}
where this rate is per galaxy ($5$ BBHs per GC, and $200$ GCs per galaxy) derived for solar type stars ($1M_{\odot},1R_{\odot}$) interacting with BBHs of equal mass. We have further assumed that $P_{\rm TD} = 2R_{\rm TD}/a$. This rate is on the lower side compared to both the derived rate of
star-SMBH TDs \citep[e.g.,][]{2014ApJ...794....9M, 2017ApJ...841..132L, 2018MNRAS.480.5060S} and long gamma ray burst (GRB)
\citep[e.g.][]{2010MNRAS.406.1944W, 2018arXiv181201118L}, but not unreasonably low
for being observationally interesting.
However, our derived rate is the rate of disruptions and not necessarily
the rate of observable events. The observable rate
depends further on the TD luminosity ($L$) and the energy extraction mechanism, which still are unsolved problems \citep[e.g.][]{2014ApJ...783...23G, 2018arXiv180110180S}. If $L$ is set by the
mass fall back rate of the tidal stream, $\dot{M}$,
then $L \propto 1/\sqrt{m_{\rm BH}}$, which suggests that lighter BHs should
result in a higher $L$ than heavier BHs.
However, if $L$ is Eddington limited, then $L \propto m_{\rm BH}$, which
immediately leads to rates that are far below observable limits \citep{2018arXiv181201118L}.
One proposed model for how radiation is processed and escapes involves
an Eddington limited accretion disk with a relativistic launched jet \citep[e.g.][]{2009ApJ...697L..77R, 2011MNRAS.416.2102G, 2012ApJ...760..103D}. In this picture
the energy flux carried by the jet can easily exceed the Eddington limit and will likely scale as
$\propto \dot{M}c^{2}$, which would make such events visible at cosmological distances.
However, in this case the observer has to be located near the cone of the jet opening,
which significantly reduces the probability for detection. Taking this effect into account,
the observed rate is likely $1/50-1/100$ times smaller \citep[e.g.][]{2010MNRAS.406.1944W, 2017ApJ...849L..29K} than the
disruption rate given by Eq. \eqref{eq:GTDEBBH}. However,
the jet might cover a larger area on the sky if precession and movement from the
BBH orbital motion is significant, which would lead to a higher probability for detection \citep[e.g.][]{2013PhRvD..87h4053S}.
All in all, compared to long GRBs, which have been observed extensively by {\it Swift}, our derived rate of star-BBH TDs from GCs is about two-orders of magnitude lower. The rate could be higher
if a subpopulation of the SCs have central stellar densities that are $> {10^{4}\text{pc}^{-3}}$, which could be the case for NSCs and GCs without a dominating BH core \citep[e.g.,][]{Kremer2018d}. The average number of GCs per galaxy could also be $> 200$. Furthermore, active star formation in NSCs leads to a population of stars with large radii which also would lead to enhanced TD rates.

Finally, if both star-BH and star-BBH TDs are observed, then their number
ratio can be used to constrain the fraction of BBHs in GCs, which play a key role in determining the BBH merger rate observable by LIGO/Virgo. To see how,
we start by writing the rate of star-BBH TDs from a
single GC as $\Gamma_{\rm TD}^{\rm BBH} \propto N_{\rm BBH} \times 4 \eta_{\rm s}m_{\rm BH}R_{\rm TD}/v_{\rm dis}$,
which follows from the arguments given in Section \ref{sec:Rate of Tidal Disruptions}, and where we have assumed that
$P_{\rm TD} = 2R_{\rm TD}/a$. Similarly, the total rate of star-BH TDs
can be written as $\Gamma_{\rm TD}^{\rm BH} \propto N_{\rm BH} \times \eta_{\rm s}m_{\rm BH}R_{\rm TD}/v_{\rm dis}$ \citep{2016ApJ...823..113P}. The ratio between
the two rates is therefore given by $\Gamma_{\rm TD}^{\rm BBH}/\Gamma_{\rm TD}^{\rm BH} \approx 4N_{\rm BBH}/N_{\rm BH} = 4f_{\rm BBH}$, where $f_{\rm BBH}$ is the number of
BBHs compared to single BHs. There are many uncertainties here,
such as single BHs and BBHs do not necessarily cluster in the same way throughout the core region, and not all star-BBH TDs will be observationally different from star-BH TDs,
but our simple relation does suggests that important astrophysical properties may be extracted from BH and BBH stellar TDs.

\subsubsection{Alternative Probes}

Other ways of observing the BBH orbital distribution includes micro-lensing \citep[e.g.,][]{2002ApJ...579..639B, 2015ApJ...810L..20M, 2016MNRAS.458.3012W},
periodic self-lensing if the BBHs accrete matter \citep[e.g.,][]{2018MNRAS.474.2975D, DOrazDiStef_InPrep}, and GW emission \citep[e.g.][]{2018arXiv180708864S, 2018arXiv181111812K}.
The possibility for detecting accreting BBHs through self-lensing is especially interesting, as this allows one
to observe the remnant system of
a star-BBH TD over a timespan comparable to the life time of the accretion disks, which is orders of magnitude
longer than the TD flares and jets we
have discussed so far. The probability for observing such systems is proportional to their life time, and
accreting BBHs will therefore in principle
have a much higher probability for being detected; however, for self-lensing to be efficient
one has to observe near the orbital plane of the BBH,
which puts strong constrains on the geometry in a similar way as for the jetted TDs.
It is not yet clear if such requirements will
reduce the number of sources below observable limits.
Other approaches for indirectly probing a possible central population of BHs is to look for correlations with the
GC stellar properties \citep[e.g.,][]{2018MNRAS.478.1844A,Kremer2018b,Kremer2018d,Weatherford2018}, for which machine learning algorithms might be particular
useful \citep[e.g.,][]{2018arXiv181106473A}.

\section{Conclusions}\label{sec:Conclusions}

The orbital time distribution of BBHs in SCs is tightly linked to the GR dynamics
that also drive the BBHs to merger. Therefore, if the BBH orbital distribution can be probed, then one will
in principle be able to indirectly constrain the current BBH merger rate from SCs observable by LIGO/Virgo,
as well as the dynamical environment of the BBHs. In this paper we propose that the BBH orbital time distribution
might be indirectly probed using interrupted stellar TDs \citep[e.g.,][]{Liu:2009fl}. The main idea behind this `multi-messenger' observable is that light-curves from star-BBH TDs are expected to show interruptions
after a time-scale comparable to the BBH orbital time due to deflections of the tidal stream by the BHs.

Using a novel analytical framework for describing the dynamical evolution of BBHs in SCs, including the possibility for GW merger during hardening, we have shown that the corresponding BBH orbital time distribution have a near universal shape that peaks at a time-scale of days (Section \ref{sec:Results}).
The distribution differs significantly from other astrophysical scenarios not driven by dynamics, including the Opik's Law distribution (flat in $\log T$), purely GW driven BBHs (increasing with $\log T$),
and SMBBH TDEs (distribute at time-scales of $10^{2}$ days).
This suggests that if interrupted TDs are observed with breaks in their light-curves appearing
after a time of hours to days (the time of interruption is usually found to be $\ll T$ \citep{Liu:2009fl}), then this would indicate a
nearby dynamical driven population of BBHs embedded in a halo of stars, such as a GC. Actively searching for
interrupted TD light-curves in current and upcoming EM surveys might therefore help us indirectly constrain the dynamical
formation of BBH mergers in nearby SCs. In addition, the fraction of BBHs to single BHs can also be determined
by probing the relative rate of star-BBH to star-BH TDs (Section \ref{sec:Discussion}).

We acknowledge that there are several uncertainties related to our proposed method, where the greatest is likely
related to the TD luminosity and thereby the observable rate. As discussed in Section \ref{sec:Discussion},
super Eddington accretion is clearly needed to create rates that are within observable limits \citep[see also:][]{2016ApJ...823..113P, 2018arXiv181201118L}; however,
exactly how a TD is powered even for single BH TDs is still an open question \citep[e.g.,][]{2015ApJ...806..164P}.
Despite this, the encouraging results from both the SMBBH limit \citep[e.g.,][]{2014ApJ...786..103L} and the stellar mass BBH limit \citep{2018arXiv181201118L} make our proposed method an interesting
possibility, and we do hope that our work
will inspire groups to perform complementary studies of
both the rate of star-BBH interactions in dense SCs as
well of the accretion process itself.

Finally, this paper also presents the first GR analytical
solution to the orbital time distribution of BBHs in dense SCs (Section \ref{sec:Distribution of BBH-Stellar Disruptions}). We have
compared our analytical solution to both a semi-analytical MC
approach (Section \ref{sec:Distribution of BBH-Stellar Disruptions}) and to
data extracted from the H\'{e}non-type MC code \texttt{CMC} (Section \ref{sec:comp_MC}).
We find excellent agreement.
As shown in Section \ref{sec:Results}, our inclusion of GR effects (representing the possibility for GW merger
inside the cluster before ejection) gives rise to significant changes compared to the purely Newtonian limit (Fig. \ref{fig:dGdlogT}). This clearly
illustrates the need for GR numerical methods for describing the distribution of BBHs even in classical GC systems \citep[e.g.,][]{2018arXiv181104926R}. We hope
our derived distribution will turn out to be useful for other studies related to BBHs in dense SCs and their detectability.

{\it Acknowledgments. ---}
The authors thank  Carl L. Rodriguez, S. Schrøder, T. Fragos, B. Mockler, S. I.
Mandel, W. Farr, C. Miller, D. J. D’Orazio, K. Hotokezaka,
M. Gaspari and A. Askar for stimulating discussions. 
JS acknowledges support
from the Lyman Spitzer Fellowship. TV acknowledges the support of the Friends of the Institute for Advanced Study.
MLJR acknowledges that all praise and thanks belongs to Allah (any
benefit is due to God and any shortcomings are my own).
ERR acknowledge support from the DNRF (Niels Bohr Professor)
and NSF grant AST-1615881. KK acknowledges support by the National Science Foundation Graduate Research Fellowship Program under Grant No. DGE-1324585. The authors further
thanks the Niels Bohr Institute for its hospitality while part
of this work was completed, and the Kavli Foundation and
the DNRF for supporting the 2017 Kavli Summer Program.

\bibliographystyle{h-physrev.bst}
\bibliography{NbodyTides_papers.bib,references.bib}

\end{document}